\documentstyle[amssymb,preprint,aps,epsfig]{revtex}
\begin{document}
\title{Phase dynamics of a multimode Bose condensate controlled by decay}
\author{H. L. Haroutyunyan and G. Nienhuis}
\address{Huygens Laboratorium, Universiteit Leiden,\\
Postbus 9504, \\
2300 RA Leiden, The Netherlands}
\maketitle

\begin{abstract}
The relative phase between two uncoupled BE condensates tends to attain a
specific value when the phase is measured. This can be done by observing
their decay products in interference. We discuss exactly solvable models for
this process in cases where competing observation channels drive the phases
to different sets of values. We treat the case of two modes which both emit
into the input ports of two beam splitters, and of a linear or circular
chain of modes. In these latter cases, the transitivity of relative phase
becomes an issue.
\end{abstract}


\section{Introduction}

Since the first observation of Bose-Einstein condensation, the formation and
the nature of the relative phase between two condensates has been a central
issue of many theoretical and experimental studies. It has been predicted by
Javanainen and Yoo \cite{Javanainen} and observed by Andrews et al \cite
{Andrews} that two interfering Bose condensates exhibit a clear spatial
interference pattern. This shows that in a single run of an interference
experiment, they manifest themselves as being coherent. Furthermore, it was
predicted in \cite{Javanainen} that two cases should be distinguished. When
a cold cloud of atoms is first split into two modes, which are separately
cooled further into two condensates ( ''cut - then- cool''), two independent
condensates arise. Alternatively, two correlated condensates arise when a
single condensate is split into two parts ('' cool - then - cut'') \cite
{Andrews,Hall}. Interference pattern from two independent condensates can be
different for each realization of interference experiment, while correlated
condensates show the same interference pattern for each run. Cirac et al 
\cite{Cirac} showed by analytical arguments that a system consisting of two
independent Bose condensates evolves into a state with a fixed relative
phase if one detects the emitted bosonic atoms while observing their spatial
interference pattern.

A number of authors have studied the possible manipulation of phase
coherence and entanglement between two or more Bose condensates, with
tunneling interaction as the key mechanism \cite
{Ruostekoski,Dunningham2,Hines}. A scheme has been proposed to use an
interferometric scheme including an atomic beam splitter to recombine two
modes in order to reconstruct the state of a two-mode condensate \cite{Bolda}%
. The buildup of a relative phase between two independent condensates has
also been investigated in the situation that the atoms emitted from the two
condensates are mixed in a 50:50 beam splitter\cite{Castin,Nienhuis}. Two
initially independent bosonic modes, described by a factorized state, evolve
into an entangled state of the two modes after a large number of detections
in the output ports of the beam splitter. The relative phase distribution
shows two narrow peaks, at positions determined by the settings of the beam
splitter. The most probable detection history has the form of bosons
bunching into a single output channel. An exactly solvable analytical model
has been discussed \cite{Nienhuis}, which allows one to get closed
expression for the particle detection statistics over two output channels of
the beam splitter for a fixed total number of detections. It is remarkable
that even though both detection channels are identical, in the most probable
history all particles are detected in the same port. This is obviously
connected to the bosonic nature of the particles, for which boson
accumulation applies. This can likewise be interpreted as a spontaneous
selection of a single relative phase$.$ When the first particle chooses
randomly one of the two output ports, the following particles have a
tendency to choose the same port, and the relative phase of the modes
converges to one of the phases imposed by the beam splitter. This can also
be viewed as an example of spontaneous symmetry breaking \cite{Sols}. The
role of interparticle interaction is also discussed, and it has been shown
that it leads to collapse and revival of the relative phase distribution,
thereby reflecting the discrete nature of the states of the system\cite
{Castin}.

In the presence of a single beam splitter, the relative phase converges
eventually to a single value. It is interesting to consider cases where more
detection channels are present which tend to project the relative phase on
different values, so that a detection from one beam splitter favors phase
values that are incompatible with the setting of another one. In the present
paper we consider a number of model cases where such a conflicting tendency
arises. This raises the question whether in the end the system simply
settles down in one of the possible phase values, or whether it continues to
shift between values, without ever coming to a final decision. We consider
cases where the detection statistics can be solved analytically. Also we
study the effect of a direct Hamiltonian coupling between the condensates on
both the detection statistics and the corresponding behavior of the relative
phase. Examples of such couplings are tunneling between condensates in two
spatially separated potential wells, or stimulated Raman coupling between
two condensates corresponding to two different internal states \cite
{Dunningham}. We treat the condensates just as modes of bosonic particles,
so that most of the considerations hold just as well for photons in cavities.

\section{Quantum states of two boson modes}

It will be convenient to express the states of two boson modes in terms of
spin-coherent states (SCS), which is normally defined for the $2J+1$%
-dimensional manifold of states with angular momentum $J$ \cite{Arecchi}.
The spin-coherent state $\left| \theta ,\phi \right\rangle $ is the
eigenstate of the component $\overrightarrow{u}\cdot \widehat{%
\overrightarrow{J}}$ of the angular momentum vector with the maximal
eigenvalue $J$, where $\overrightarrow{u}\equiv \widehat{x}\cos \phi \sin
\theta +\widehat{y}\sin \phi \sin \theta +\widehat{z}\cos \theta $ is the
unit vector in the direction specified by the spherical angles $\theta $ and 
$\phi $. This state is obtained from the eigenstate of $J_{z}$ with
eigenvalue $J$ after performing the appropriate rotation. In the context of
two boson modes (or two harmonic oscillators), an $SU(2)$ representation
arises by introducing the fictitious angular-momentum operators 
\begin{equation}
\widehat{J}_{x}={\frac{1}{2}}\left( \widehat{a}^{\dagger }\widehat{b}+%
\widehat{b}^{\dagger }\widehat{a}\right) ,\;\widehat{J}_{y}=\frac{1}{2i}%
\left( \widehat{a}^{\dagger }\widehat{b}-\widehat{b}^{\dagger }\widehat{a}%
\right) ,\widehat{J}_{z}={\frac{1}{2}}\left( \widehat{a}^{\dagger }\widehat{a%
}-\widehat{b}^{\dagger }\widehat{b}\right) ,  \label{Schwinger}
\end{equation}
where $\widehat{a}$ and $\widehat{b}$ are the annihilation operators for
modes $A$ and $B$. This is the well-known Schwinger representation. These
operators obey the standard commutation rules of angular momentum ($[%
\widehat{J}_{x},\widehat{J}_{y}]=i\widehat{J}_{z}$, etc.), so that the
matrix form of the operators (\ref{Schwinger}) on the eigenvectors of $%
\widehat{J}_{z}$ and $\widehat{\overrightarrow{J}}^{2}$ attains the shape
that is well-known from angular-momentum algebra. Notice that $\widehat{%
\overrightarrow{J}}^{2}=\frac{\widehat{N}}{2}(\frac{\widehat{N}}{2}+1)$,
with $\widehat{N}=\widehat{a}^{\dagger }\widehat{a}+\widehat{b}^{\dagger }%
\widehat{b}$ the number operator. The eigenvectors of $\widehat{J}_{z}$ and $%
\widehat{\overrightarrow{J}}^{2}$are just the double Fock states $\left|
n_{a},n_{b}\right\rangle $. A given number of particles $N$ corresponds to
the value $J=N/2$. The eigenstate of $\widehat{J}_{z}$ with this same
eigenvalue is the Fock state $\left| N,0\right\rangle $, so that the SCS
with direction $\overrightarrow{u}$can be defined by the rotation 
\begin{equation}
\left| \theta ,\phi \right\rangle _{N}=\widehat{R}(\theta ,\phi )\left|
N,0\right\rangle ,  \label{SCS}
\end{equation}
with the rotation operator 
\begin{equation}
\widehat{R}(\theta ,\phi )=\exp (-i\phi \widehat{J}_{z})\exp (-i\theta 
\widehat{J}_{y})\exp (i\phi \widehat{J}_{z})=\exp [-i\theta (\widehat{J}%
_{y}\cos \phi -\widehat{J}_{x}\sin \phi )].  \label{rotation}
\end{equation}
The SCS can be represented as a point on a sphere of radius $J$, specified
by the polar angle $\theta $, and the azimuthal angle $\phi $. This sphere
generalizes the Bloch sphere, describing the state of a spin $1/2$, or the
Poincar\'{e} sphere which describes the polarization state of a light beam
or a photon. In the present case, the radius specifies the number of
particles $N=2J$. An explicit expansion of the SCS (\ref{SCS}) in the Fock
states follows then from the transformation of the creation operators 
\begin{equation}
\widehat{R}(\theta ,\phi )\widehat{a}^{\dagger }\widehat{R}^{\dagger
}(\theta ,\phi )=\widehat{a}^{\dagger }\cos \frac{\theta }{2}+\widehat{b}%
^{\dagger }\sin \frac{\theta }{2}e^{i\phi }\equiv \widehat{c}^{\dagger
}(\theta ,\phi ).  \label{trans}
\end{equation}
The SCS (\ref{SCS}) is found after operating $N$ times with the operator $%
\widehat{c}^{\dagger }(\theta ,\phi )$ on the vacuum state, which leads to
the explicit result 
\begin{equation}
\left| \theta ,\phi \right\rangle _{N}=\sum_{n=0}^{N}\left( 
\begin{array}{l}
N \\ 
n
\end{array}
\right) ^{1/2}\cos ^{n}\frac{\theta }{2}\text{ }\sin ^{N-n}\frac{\theta }{2}%
\text{ }e^{i\left( N-n\right) \phi }\text{ }\left| n,N-n\right\rangle .
\label{SCSexp}
\end{equation}
This demonstrates that the SCS $\left| \theta ,\phi \right\rangle _{N}$ can
be viewed as a number state in the mode that is a linear combination of the
modes $A$ and $B$, and for which the operator $\widehat{c}^{\dagger }(\theta
,\phi )$, defined in (\ref{trans}), is the creation operator. In the SCS,
the distribution of the $N$ particles over the two modes is binomial, and
the angle $\theta $ specifies the average partition by $\left\langle
n_{a}\right\rangle =N\cos ^{2}\frac{\theta }{2}$ and $\left\langle
n_{b}\right\rangle =N\sin ^{2}\frac{\theta }{2}$. The azimuthal angle $\phi $
represents the relative phase. This quantity is complementary to the number
difference $\widehat{a}^{\dagger }\widehat{a}-\widehat{b}^{\dagger }\widehat{%
b}$. Number states with all particles in the mode $A$ are represented by the
North pole of the Bloch sphere ($\theta =0$), while the South pole
represents the SCS with all $N$ particles in mode $B$. Points on the equator
($\theta =\pi /2$) stand for states with equal population of the modes.
Since the state (\ref{SCS}) (or (\ref{SCSexp})) is eigenstate of $\widehat{N}
$, the absolute phase is fully undetermined.

The relation between the SCS and the more common Glauber coherent states
(GCS) is easily found by representing the latter ones in the form 
\begin{equation}
\left| r_{a}e^{-i\phi _{a}},r_{b}e^{-i\phi _{b}}\right\rangle
=e^{-(r_{a}^{2}+r_{b}^{2})/2}\sum_{N}\frac{1}{N!}(r_{a}e^{-i\phi _{a}}%
\widehat{a}^{\dagger }+r_{b}e^{-i\phi _{b}}\widehat{b}^{\dagger })^{N}\left|
vac\right\rangle .  \label{GCS}
\end{equation}
These states are eigenstates of $\widehat{a}$ and $\widehat{b}$, and they
are obviously factorized, so that they carry no entanglement between the
modes. It is easy to check that they are related to the SCS by the expansion 
\begin{equation}
\left| r_{a}e^{-i\phi _{a}},r_{b}e^{-i\phi _{b}}\right\rangle
=e^{-R^{2}/2}\sum_{N}\frac{1}{\sqrt{N!}}R^{N}e^{-iN\phi _{a}}\left| \theta
,\phi \right\rangle _{N},  \label{S-GCS}
\end{equation}
with the parameters $R$, $\theta $ and $\phi $ determined by $%
R^{2}=r_{a}^{2}+r_{b}^{2}$, $\tan \frac{\theta }{2}=r_{b}/r_{a}$, and $\phi
=\phi _{a}-\phi _{b}$. This indicates that the GCS has a Poissonian
distribution of the total particle number $N$, with average value $%
\left\langle N\right\rangle =R^{2}$, while the absolute phases $\phi _{a}$
and $\phi _{b}$ of both modes are well-specified. For bosonic atoms, states
with different total number of particles do not superpose, according to the
superselection rule, so that we have to restrict ourselves to density
matrices that are diagonal in $N$. Since the particle number is conjugate to
the overall phase, we introduce the density matrix 
\begin{equation}
\widehat{\rho }(R,\theta ,\phi )=\frac{1}{2\pi }\int_{0}^{2\pi }d\phi
_{a}\left| r_{a}e^{-i\phi _{a}},r_{b}e^{-i(\phi _{a}-\phi )}\right\rangle
\left\langle r_{a}e^{-i\phi _{a}},r_{b}e^{-i(\phi _{a}-\phi )}\right|
\label{rhoGCS}
\end{equation}
as the uniform mixture of the GCS (\ref{GCS}) over the overall phase $\phi
_{a}$, for a given value of the relative phase $\phi =\phi _{a}-\phi _{b}$.
Applying eq. (\ref{S-GCS}) leads to an expansion of this same density matrix
in the SCS, in the form 
\begin{equation}
\widehat{\rho }(R,\theta ,\phi )=e^{-R^{2}}\sum_{N}\frac{1}{N!}R^{2N}\left|
\theta ,\phi \right\rangle _{N}{}_{N}\left\langle \theta ,\phi \right| .
\label{rhoSCS}
\end{equation}
The density matrix $\widehat{\rho }(R,\theta ,\phi )$ is therefore diagonal
in the particle number $N$.

We observe that to each pair of spherical angles $\theta $ and $\phi $, or,
equivalently, to each real Cartesian unit vector $\overrightarrow{u}$,
corresponds a density matrix $\widehat{\rho }(R,\theta ,\phi )$, and an
annihilation operator $\widehat{c}(\theta ,\phi )$, as defined in (\ref
{trans}). Now consider the annihilation operator $\widehat{c}(\theta
_{0},\phi _{0})$, corresponding to the unit vector $\overrightarrow{u}_{0}.$

In this paper we shall use density matrices that can be represented as a
superposition of the states (\ref{rhoSCS}) for a single value of the
strength parameter $R$, in the form 
\begin{equation}
\int d\Omega f(\theta ,\phi )\widehat{\rho }(R,\theta ,\phi ),  \label{dens}
\end{equation}
where we use the abbreviation $\int d\Omega =\int_{0}^{2\pi }d\phi
\int_{0}^{\pi }d\theta \sin \theta $ for the integration over the Bloch
sphere. When we express $\widehat{\rho }(R,\theta ,\phi )$ as in eq. (\ref
{rhoGCS}), it becomes clear that eq. (\ref{inital}) is just the two-mode
version of the Glauber-Sudarshan diagonal coherent-state representation of
the initial density matrix \cite{Mandel}, where the $P$-distribution is
uniform in $\phi _{A}$, and is non-zero only for a single value of $R$. This
state is normalized as soon as the distribution $f$ is, which we shall
assume. Another special case arises when the function $f$ is nonzero only
for a single value of $\theta $, and uniform in $\phi $. Then the density
matrix (\ref{dens}) can be written as 
\begin{equation}
\int d\phi \widehat{\rho }(R,\theta ,\phi )/2\pi .  \label{densfact}
\end{equation}
It follows from the coherent-state representation (\ref{rhoGCS}) that in
this case the density matrix factorizes into a product of separate density
matrices for the two modes, implying that the state (\ref{densfact}) not
entangled. The phase of both modes is uniformly distributed, and the state
is diagonal in both particle numbers $n_{a}$ and $n_{b}$.

\section{Decay and detection statistics of two boson modes}

\subsection{Master equation and detection histories}

We assume that particles are leaking out of the two boson modes $A$ and $B$,
at a total loss rate $\Gamma $. The emitted particles are detected after
passing through a beam splitter. For simplicity, we assume perfect detection
efficiency. Moreover, their evolution is governed by a Hamiltonian $\widehat{%
H}$ that is supposed to commute with the number operator $\widehat{N}$, and
which describes the energy per particle, and possibly interparticle
interaction or tunneling between the modes. Since the two modes form an open
system, their evolution can be described by a quantum master equation \cite
{Gardiner,Mandel} for the two-mode density matrix $\widehat{\rho }$, which
we formally express as 
\begin{equation}
\frac{d\widehat{\rho }}{dt}\equiv \left( {\cal L}_{0}+{\cal L}_{1}\right) 
\widehat{\rho }.  \label{master}
\end{equation}
Here ${\cal L}_{0}$ describes the coherent evolution of the system, which is
determined by the Hamiltonian evolution, and the loss of the probability of
states due to the emission of particles. Its explicit form is given by its
action on a density matrix 
\begin{equation}
{\cal L}_{0}\widehat{\rho }=-\frac{i}{\hbar }\left[ \widehat{H},\widehat{%
\rho }\right] -\frac{1}{2}\Gamma \left( \widehat{N}\widehat{\rho }+\widehat{%
\rho }\widehat{N}\right) ,  \label{L0}
\end{equation}
while the compensating probability gain is accounted for by 
\begin{equation}
{\cal L}_{1}\widehat{\rho }=\Gamma \left( \widehat{a}\widehat{\rho }\widehat{%
a}^{\dagger }+\widehat{b}\widehat{\rho }\widehat{b}^{\dagger }\right) .
\label{L1}
\end{equation}
For simplicity the loss rate of the two modes is taken to be the same. The
solution of (\ref{master}) describes the evolution of the system averaged
over all possible detection histories. In fact, we are interested in the
conditional evolution for specific histories, where the arrival times for
particles at each detector are specified. Depending on the specific setup,
we have to separate the total gain term (\ref{L1}) in terms corresponding to
each detector separately, in accordance with the method of quantum
trajectories \cite{Castin,Cirac,Nienhuis}. For instance, when a detector is
directly coupled to each mode, the term $\widehat{a}\widehat{\rho }\widehat{a%
}^{\dagger }$ describes the effect of a detection of a particle from mode $A$%
, which corresponds to the annihilation of a particle from this mode. Now we
consider the setup sketched in Fig. 1, where each mode emits particles into
the input port of two different beam splitters. Detections in the two output
ports of beam splitter I correspond to the detection operators $\widehat{c}%
_{\pm }=(\widehat{a}\pm \widehat{b})/\sqrt{2}$, and detections in the output
ports of beam splitter II correspond to the detection operators $\widehat{d}%
_{\pm }=(\widehat{a}\pm e^{-i\xi }\widehat{b})/\sqrt{2}$. The relative
phases can be set either by using dephasers, or by differences in the
pathlengths of the channels. Notice that the detection operators are
annihilation operators corresponding to a spin-coherent state that is
represented by points on the equator of the Bloch sphere. For this setup the
gain operator ${\cal L}_{1}$ can be separated into four terms corresponding
to the four detectors as 
\begin{equation}
{\cal L}_{1}\widehat{\rho }=\frac{\Gamma }{2}\left( \widehat{c}_{+}\widehat{%
\rho }\widehat{c}_{+}^{\dagger }+\widehat{c}_{-}\widehat{\rho }\widehat{c}%
_{-}^{\dagger }+\widehat{d}_{+}\widehat{\rho }\widehat{d}_{+}^{\dagger }+%
\widehat{d}_{-}\widehat{\rho }\widehat{d}_{-}^{\dagger }\right) \equiv \frac{%
\Gamma }{2}\sum_{s=1}^{4}\widehat{c}_{s}\widehat{\rho }\widehat{c}%
_{s}^{\dagger }=\sum_{s=1}^{4}{\cal L}_{1s}\widehat{\rho }.  \label{L1sep}
\end{equation}
The integral form of the master equation (\ref{master}) 
\begin{equation}
\widehat{\rho }\left( T\right) =e^{{\cal L}_{0}T}\text{ }\widehat{\rho }%
\left( 0\right) +\sum_{i}\int_{0}^{T}dt\text{ }e^{{\cal L}_{0}\left(
T-t\right) }{\cal L}_{1i}\widehat{\rho }\left( t\right)
\label{integmastereq}
\end{equation}
allows us after iteration to express the density matrix as a summation and
integration over detection histories. The contribution to $\widehat{\rho }%
\left( T\right) $ from the history with detections at the successive time
instants $t_{1}\leq t_{2}\leq \ldots \leq t_{L}$ by the detectors $s_{1}$, $%
s_{2}$, $\ldots s_{L}$ in the time interval $\left[ 0,T\right] $ is
described by the operator 
\begin{equation}
\widehat{\rho }_{L}\left( \left\{ t_{i},s_{i}\right\} ,T\right) =e^{{\cal L}%
_{0}\left( T-t_{L-1}\right) }{\cal L}_{1s_{L}}e^{{\cal L}_{0}\left(
t_{L}-t_{L-1}\right) }\ldots {\cal L}_{1s_{1}}e^{{\cal L}_{0}t_{1}}\widehat{%
\rho }\left( 0\right) .  \label{integmaster}
\end{equation}
The effect of the detection operators ${\cal L}_{1i}$ is a sudden change in
the density matrix, which indicates the quantum-jump nature of a detection.

\subsection{Detection statistics and phase distribution}

As initial state $\widehat{\rho }(0)$ of the system we take a density matrix
of the form (\ref{dens}), so that 
\begin{equation}
\widehat{\rho }(0)=\int d\Omega f(\theta ,\phi )\widehat{\rho }(R,\theta
,\phi ).  \label{inital}
\end{equation}
When the Hamiltonian only attributes a fixed energy per particle, its form
is $\widehat{H}=\hbar \omega \widehat{N}$. Since all density matrices that
we shall encounter are diagonal in the total number of particles, the
Hamiltonian has no effect, and can be ignored. The coherent evolution of the
density matrix is easily obtained from the identity ${\cal L}_{0}\left| \phi
,\theta \right\rangle _{NN}\left\langle \theta ,\phi \right| =-\Gamma
N\left| \phi ,\theta \right\rangle _{NN}\left\langle \theta ,\phi \right| $,
which when substituted into eq. (\ref{rhoSCS}) gives the result 
\begin{equation}
e^{{\cal L}_{0}T}\widehat{\rho }(R,\theta ,\phi )=\exp [-R^{2}(1-e^{-\Gamma
T})]\widehat{\rho }(Re^{-\Gamma T/2},\theta ,\phi ).  \label{cohevol}
\end{equation}
This shows that the evolution of the density matrix during a detection-free
period of time only gives a damping of the strength parameter $R$, without
changing the distribution over the Bloch sphere. The action of the detection
operators on the density matrix is most easily obtained by using eq. (\ref
{rhoGCS}). The action of the annihilation operators on the SCS is found to
be given by 
\begin{equation}
\widehat{a}\left| \theta ,\phi \right\rangle _{N}=\sqrt{N}\cos \frac{\theta 
}{2}\left| \theta ,\phi \right\rangle _{N-1},\widehat{b}\left| \theta ,\phi
\right\rangle _{N}=\sqrt{N}\sin \frac{\theta }{2}e^{i\phi }\left| \theta
,\phi \right\rangle _{N-1}  \label{aSCS}
\end{equation}
Then a direct calculation shows that 
\begin{equation}
\widehat{c}(\theta _{0},\phi _{0})\widehat{\rho }(R,\theta ,\phi )\widehat{c}%
^{\dagger }(\theta _{0},\phi _{0})=\frac{1}{2}R^{2}(1+\overrightarrow{u}%
\cdot \overrightarrow{u}_{0})\widehat{\rho }(R,\theta ,\phi ),  \label{jump}
\end{equation}
with $\widehat{c}$ defined in eq. (\ref{trans}). The unit vectors $%
\overrightarrow{u}$ and $\overrightarrow{u}_{0}$ in eq. (\ref{jump}) are
defined to point in the directions specified by the angles ($\theta ,\phi )$
and ($\theta _{0},\phi _{0})$ respectively. This indicates that for these
operators $\widehat{c}\widehat{\rho }\widehat{c}^{\dagger }$ is proportional
to $\widehat{\rho }.$ The proportionality factor takes the maximal value $%
R^{2}$ when the two directions $\overrightarrow{u}_{0}$ and $\overrightarrow{%
u}$ coincide, and it is zero when the directions are opposite. It is not
surprising that this factor depends only on the inner product of the two
unit vectors, and thereby on the distance between the two points on the unit
sphere. Application of (\ref{jump}) leads to the expression 
\begin{equation}
{\cal L}_{1s}\widehat{\rho }(R,\theta ,\phi )=\Gamma R^{2}g_{s}(\theta ,\phi
)\widehat{\rho }(R,\theta ,\phi ),  \label{jumps}
\end{equation}
where the functions $g_{i}$ for the detectors $1$ and $2$ are given by 
\begin{equation}
g_{1}(\theta ,\phi )=\frac{1}{4}(1+\sin \theta \cos \phi ),g_{2}(\theta
,\phi )=\frac{1}{4}(1-\sin \theta \cos \phi ),  \label{g12}
\end{equation}
and for the detectors $3$ and $4$ by 
\begin{equation}
g_{3}(\theta ,\phi )=\frac{1}{4}(1+\sin \theta \cos (\phi -\xi
)),g_{4}(\theta ,\phi )=\frac{1}{4}(1-\sin \theta \cos (\phi -\xi )).
\label{g34}
\end{equation}
The functions are determined by the inner product of the unit vector $%
\overrightarrow{u}$, indicated by $\theta $ and $\phi $, and the unit
vectors $\overrightarrow{u}_{0}$ corresponding to the detection operators $%
\widehat{c}_{s}$. These four unit vectors are all defined by $\theta
_{0}=\pi /2$, whereas $\phi _{0}=0$ and $\pi $ for $s=1$ and $2$, and $\phi
_{0}=\xi $ and $\xi +\pi $ for $s=3$ and $4$. The functions $g_{s}$ add up
to $1$, so that the total gain operator ${\cal L}_{1}$ when acting on $%
\widehat{\rho }(R,\theta ,\phi )$ just gives the factor $\Gamma R^{2}$, as
it should. According to eq. (\ref{jumps}), the effect of the $i$th detection
at time $t_{i}$ by detector $s_{i}$ is that the distribution over the Bloch
sphere is multiplied by the factor $g_{s_{i}}$, while an overall factor $%
\Gamma R^{2}\exp (-\Gamma t_{i})$ has to be added. In brief, the
detection-free periods produce a damping of $R$, and the detection modify
the distribution over the Bloch sphere by a multiplication with a function $%
g_{s_{i}}$. For a given value of the ratio $\left\langle n_{a}\right\rangle
/\left\langle n_{b}\right\rangle $, as specified by the angle $\theta $, the
factors $g_{s}$ modify the distribution over the relative phase $\phi $,
with a contrast that is maximal when both modes contain the same number of
particles ($\theta =\pi /2$).

The eqs. (\ref{cohevol})-(\ref{g34}) allow one to evaluate explicitly the
density matrix (\ref{integmaster}) corresponding to a given detection
history, with the initial state determined by (\ref{inital}). The
contribution (\ref{integmaster}) to the density matrix is then found as 
\[
\widehat{\rho }_{L}\left( \left\{ t_{i},s_{i}\right\} ,T\right) =\exp
[-R^{2}(1-e^{-\Gamma T})]\prod_{i=1}^{L}(\Gamma R^{2}e^{-\Gamma t_{i}}) 
\]
\begin{equation}
\times \int d\Omega f(\theta ,\phi )\left[
\prod_{s=1}^{4}g_{s}^{n_{s}}(\theta ,\phi )\right] \widehat{\rho }%
(Re^{-\Gamma T/2},\theta ,\phi ),  \label{partialdens}
\end{equation}
with $n_{s}$ the total number of detections in channel $s$ (with $\sum
n_{s}=L$). This contribution (\ref{partialdens}) does not depend on the
specific order of the detections in the various channels. The trace of (\ref
{partialdens}) specifies the probability distribution of the detection
history $\left\{ t_{i},s_{i}\right\} $ in the factorized form 
\begin{equation}
w_{L}\left( \left\{ t_{i},s_{i}\right\} ,T\right) =F(\{n_{s}\})\exp
[-R^{2}(1-e^{-\Gamma T})]\prod_{i=1}^{L}(\Gamma R^{2}e^{-\Gamma t_{i}}),
\label{probdens}
\end{equation}
with 
\begin{equation}
F(\{n_{s}\})=\int d\Omega f(\theta ,\phi
)\prod_{s=1}^{4}g_{s}^{n_{s}}(\theta ,\phi )  \label{FL}
\end{equation}
the probability that $L$ successive detections occur in the specific order ($%
s_{1}$, $s_{2}$,$\ldots $,$s_{L}$). This factor $F$ only depends on the
number of detections $n_{s}$ for each channel, not on the time ordering of
the detections. The remaining time-dependent factor in (\ref{probdens}) is
the probability density for detections at the specified instants of time,
irrespective of the detection channel. The conditional density of the
system, given the detection history $\left\{ t_{i},s_{i}\right\} $, is equal
to $\widehat{\rho }_{L}\left( \left\{ t_{i},s_{i}\right\} ,T\right)
/w_{L}\left( \left\{ t_{i},s_{i}\right\} ,T\right) $, which is the
normalized version of (\ref{partialdens}). From the expression (\ref
{probdens}) of the probability density one obtains the probability $%
p(\{n_{s}\},T)$ that in the time interval $[0,T]$ there were $n_{s}$
detections in channel $s$, ($s=1$,$\ldots $,$4$), irrespective of the order
of the detections. This requires an integration over the ordered detection
times, and a multiplication with the number of possible orderings of the $L$
detections over the four detectors, given the partition $\{n_{s}\}$. The
result can be expressed as 
\begin{equation}
p(\{n_{s}\},T)=P_{L}(T)p_{L}(\{n_{s}\}),  \label{pT}
\end{equation}
where $P_{L}(T)$ gives the probability that precisely $L$ detections
occurred in the time interval $[0,T]$, irrespective of the detection
channel. This distribution is Poissonian with average $R^{2}(1-e^{-\Gamma
T}) $. The factor $p_{L}(\{n_{s}\})$ is the probability that the $L$
detections are distributed over the four detectors by the partition $%
\{n_{s}\}$, and takes the form 
\begin{equation}
p_{L}(\{n_{s}\})=\frac{L!}{n_{1}!n_{2}!n_{3}!n_{4}!}F(\{n_{s}\}).  \label{pL}
\end{equation}
This distribution is independent of the strength factor $R$, the detection
time $T$ and the decay rate $\Gamma $. Notice that both the distribution $%
P_{L}(T)$ over the total number $L$ of detections, and the distribution $%
p_{L}(\{n_{s}\})$ of the $L$ detections over the partitions are normalized.

In summary, we notice that the decay process only has the effect that the
strength factor $R$ is damped. The effect of a detection is that the
distribution over the Bloch sphere is multiplied by one of the factors $%
g_{s} $, which changes both the distribution over the relative phase and the
probability distribution for subsequent detections. The probability
distribution of $L$ detections over the four detection channels is given by (%
\ref{pL}). After a detection series given by the partition $\{n_{s}\}$, the
normalized distribution function over the Bloch sphere is given by the $%
f(\theta ,\phi )\prod_{s}g_{s}^{n_{s}}(\theta ,\phi )/$ $F(\{n_{s}\})$.

\subsection{Sum rules}

When the detections in the channels $3$ and $4$ are ignored, and $M$
detections have occurred in the channels $1$ and $2$, the distribution of
these detections over the two channels can be evaluated in the same fashion.
The result is 
\begin{equation}
p_{M}(n_{1},n_{2})=2^{M}%
{M \choose n_{1}}%
\int d\Omega f(\theta ,\phi )g_{1}^{n_{1}}(\theta ,\phi
)g_{2}^{n_{2}}(\theta ,\phi ),  \label{qM}
\end{equation}
with $n_{1}+n_{2}=M$. The factor $2^{M}$ is needed to ensure normalization,
since $g_{1}+g_{2}=1/2$ in this case. This expression is a simple
generalization of the result of \cite{Nienhuis} for the case of two decaying
modes observed through a single beam splitter. The generalization consists
in the fact that the populations of the two modes need not be the same in
eq. (\ref{qM}). Intuitively it is obvious that the partial statistics of
detections in the channels $1$ and $2$ is not affected when for some reason
the detections in the channels $3$ and $4$ are simply added without
distinguishing them. This situation is equivalent to the case that beam
splitter II is missing, and a single detector is just collecting particles
in both of its input channels. For a total number $L$ of detections, the
probability of having $n_{1}$ and $n_{2}$ detections in channels $1$ and $2$%
, with $n_{1}+n_{2}=M\leq L$ can be expressed as a marginal distribution of $%
p_{L}(\{n_{s}\})$ in which the sum $n_{3}+n_{4}$ is fixed. When using that $%
g_{3}+g_{4}=1/2$ we find 
\begin{equation}
\sum_{n_{3}+n_{4}=L-M}p_{L}(\{n_{s}\})=%
{L \choose M}%
2^{-L}p_{M}(n_{1},n_{2}),  \label{marg}
\end{equation}
with $p_{M}$ the distribution (\ref{qM}) over channels $1$ and $2$,
regardless the detections in channels $3$ and $4$. Ths confirms that the
relative distribution of the detections over the first two channels remains
unaffected by the detections in the channels $3$ and $4$, provided that
these are not distinguished.

\subsection{Special cases}

We have noticed that the effect of detections on the phase distribution is
strongest when the average number of particles is the same in both modes, so
we consider the case that the polar angle is $\theta =\pi /2$, so that $%
r_{a}=r_{b}=R/\sqrt{2}\equiv r$. For this situation, the two-channel
distribution (\ref{qM}) has been evaluated in ref. \cite{Nienhuis}. When the
relative phase $\phi $ has a well-defined value $\phi _{0}$, the two-channel
distribution is binomial 
\begin{equation}
p_{M}(n_{1},n_{2})=%
{M \choose n_{1}}%
\cos ^{2n_{1}}\frac{\phi _{0}}{2}\sin ^{2n_{2}}\frac{\phi _{0}}{2},
\label{binomial}
\end{equation}
where the most probable detection history has the values $n_{1}=M\cos
^{2}(\phi _{0}/2)$, $n_{2}=M\sin ^{2}(\phi _{0}/2)$. When the phase
distribution is uniform, the two-channel distribution was found as \cite
{Nienhuis} 
\begin{equation}
p_{M}(n_{1},n_{2})=\frac{1}{2^{2M}}%
{2n_{1} \choose n_{1}}%
{2n_{2} \choose n_{2}}%
,  \label{uniform}
\end{equation}
which displays boson accumulation, with the most probable history specified
by $(n_{1},n_{2})=(M,0)$ or $(0,M)$. After such a history, the
relative-phase distribution is proportional to $\cos ^{2M}\frac{\phi }{2}$
or $\sin ^{2M}\frac{\phi }{2}$, which peaks at the positions corresponding
to the output channels of the beam splitter $I$.

Now we turn to the detection statistics over the four channels when the
initial density matrix is specified by eq. (\ref{densfact}), with equal
population of the two modes, and initial uniform relative phase. Then the
initial density matrix (\ref{inital}) is equivalent to the factorized form $%
\widehat{\rho }(0)=\widehat{\rho }_{a}\otimes \widehat{\rho }_{b}$, with 
\begin{equation}
\widehat{\rho }_{a}=\frac{1}{2\pi }\int d\phi _{a}\left| re^{-i\phi
_{a}}\right\rangle \left\langle re^{-i\phi _{a}}\right| ,  \label{poissonian}
\end{equation}
and a similar expression for $\widehat{\rho }_{b}$. Both modes have a
density matrix that is diagonal in the number state, with a Poissonian
distribution. Intuitively one would expect that both two-channel
distributions (\ref{binomial}) and (\ref{uniform}) are contained in the
margins of the four-channel distribution $p_{L}(\{n_{s}\})$, which must be
equal to the product of the marginal distribution (\ref{marg}) with $%
M=n_{1}+n_{2}$, and the conditional distribution $%
p_{L}(n_{3},n_{4}|n_{1},n_{2})$. We look for detection histories with
maximum probability. First we notice that the emission probability onto both
beam splitters I and II is the same, so that for a total of $L$ detections a
most probable history must have $n_{1}+n_{2}=n_{3}+n_{4}=L/2$. (We assume
that $L$ is even for simplicity.) If nothing is specified on the
distribution of the $L/2$ detections in the channels $3$ and $4$, the
distribution over the two channels $1$ and $2$ is given by eq. (\ref{uniform}%
) with $M=L/2$, with the most probable partitions $(n_{1},n_{2})=(L/2,0)$ or 
$(0,L/2)$. The relative phase has then converged to the value $\phi =0$ or $%
\phi =\pi $, which makes the distribution over the $L/2$ detections in
channels $3$ and $4$ binomial. For example, for the partition $%
(n_{1},n_{2})=(L/2,0)$, the partition over the two other detectors has
maximal probability for $(n_{3},n_{4})=(L/2)(\cos ^{2}(\xi /2),\sin ^{2}(\xi
/2))$. Since the pair of detectors $1$ and $2$ is fully equivalent to the
pair $3$ and $4$, another history with the same maximal probability occurs
for the partition $(n_{3},n_{4})=(L/2,0)$, with $(n_{1},n_{2})=(L/2)(\cos
^{2}(\xi /2),\sin ^{2}(\xi /2))$. This corresponds to a relative phase
converging to the value $\phi =\xi $. In summary, we expect four most
probable histories for $L$ detections. The partitions over the four
detectors attain the values $(n_{1},n_{2},n_{3},n_{4})=(L/2)(1,0,\cos
^{2}(\xi /2),\sin ^{2}(\xi /2))$, $(L/2)(0,1,\sin ^{2}(\xi /2),\cos ^{2}(\xi
/2))$, $(L/2)(\cos ^{2}(\xi /2),\sin ^{2}(\xi /2),1,0)$ and $(L/2)(\sin
^{2}(\xi /2),\cos ^{2}(\xi /2),0,1)$, while the phase has converged in these
cases to the values $\phi =0$, $\pi $, $\xi $ and $\xi +\pi $, respectively.
These considerations are backed up by a numerical calculation of the
probability distribution $p_{L}(\{n_{s}\})$, for $L=40$, equal population of
the two wells ($\theta =\pi /2$), uniform distribution over the relative
phase $\phi $, while the setting of the two beam splitters is maximally
different ($\xi =\pi /2$). The distribution for equal number of detections
through both beam splitters is plotted in Fig. 2. The most probable
histories are marked. The gradual transition between the two distributions (%
\ref{binomial}) and (\ref{uniform}) is noticed along the axis $n_{1}$, when $%
n_{3}$ varies from $0$ (binomial distribution over $n_{1}$ and $%
n_{2}=L/2-n_{1}$) and $L/2$ (accumulation distribution (\ref{uniform})).

\section{Detection statistics of two coupled boson modes}

\subsection{Pulsed coupling between modes}

In this secton, we consider the case that the particles emitted by the two
boson modes $A$ and $B$ are detected directly, without the use of beam
splitters, as sketched in Fig. 3. Therefore we separate the gain operator in
the master equation (\ref{master}) as ${\cal L}_{1}={\cal L}_{1a}+{\cal L}%
_{1b}$, corresponding to the two terms in (\ref{L1}). The coherent-evolution
operator ${\cal L}_{0}$ is given by eq. (\ref{L0}), where the Hamiltonian $%
\widehat{H}$ describes coupling between the two modes by tunneling, in the
form 
\begin{equation}
\widehat{H}=-\frac{\hbar \delta }{2}\left( \widehat{a}^{\dagger }\widehat{b}+%
\widehat{a}\widehat{b}^{\dagger }\right) =-\hbar \delta \widehat{J}_{x}.
\label{Hcoup}
\end{equation}
In realistic cases we can imagine that the coupling can be switched on
during a time interval $\tau $, which is sufficiently small so that decay
during the coupling is negligible. This means that the initial state for the
decay process is found by applying the pulse evolution operator 
\begin{equation}
\widehat{U}_{0}=\exp (-i\widehat{H}\tau /\hbar )=\exp (i\delta \tau \widehat{%
J}_{x}).  \label{pulse}
\end{equation}
In the picture of the Bloch sphere, this is a rotation about the $x$-axis in
a negative direction over an angle $\delta \tau $. When the initial state
before the coupling is given by (\ref{dens}), the state after switching-off
the coupling at the beginning of the detection period is 
\begin{equation}
\widehat{\rho }(0)=\int d\Omega f(\theta ,\phi )\widehat{U}_{0}\widehat{\rho 
}(R,\theta ,\phi )\widehat{U}_{0}^{\dagger }.  \label{initcoup}
\end{equation}
The contribution to the density matrix from a given detection history $%
\{t_{i},s_{i}\}$ is expressed by eq. (\ref{integmaster}), where now the
indices $s$ of the jump operators ${\cal L}_{1s}$ can take the values $a$ or 
$b$, and where eq. (\ref{initcoup}) specifies the initial density matrix.
The evolution during the detection-free periods is given in eq. (\ref
{cohevol}). The effect of the jump operators on the rotated density matrix
can be expressed using the identity 
\[
{\cal L}_{1a}\widehat{U}_{0}\widehat{\rho }\widehat{U}_{0}^{\dagger }=\Gamma 
\widehat{U}_{0}\widehat{c}_{a}\widehat{\rho }\widehat{c}_{a}^{\dagger }%
\widehat{U}_{0}^{\dagger }, 
\]
and a similar expression for ${\cal L}_{1b}$, where we introduced the
counterrotated operators $\widehat{c}_{a}\equiv \widehat{U}_{0}^{\dagger }%
\widehat{a}\widehat{U}_{0}$ and $\widehat{c}_{b}\equiv \widehat{U}%
_{0}^{\dagger }\widehat{b}\widehat{U}_{0}$. Their explicit expressions are
then 
\[
\widehat{c}_{a}=\widehat{a}\cos \frac{\delta \tau }{2}+i\widehat{b}\sin 
\frac{\delta \tau }{2},\widehat{c}_{b}=i\widehat{a}\sin \frac{\delta \tau }{2%
}+\widehat{b}\cos \frac{\delta \tau }{2}. 
\]
They correspond in the sense of eq. (\ref{trans}) to the two unit vectors $%
\overrightarrow{u}_{a}=-\widehat{y}\sin \delta \tau +\widehat{z}\cos \delta
\tau $, $\overrightarrow{u}_{b}=\widehat{y}\sin \delta \tau -\widehat{z}\cos
\delta \tau $, which arise when the opposite rotation is applied to $\pm 
\widehat{z}$. By using eq. (\ref{jump}), the action of the jump operators $%
{\cal L}_{1a}$ and ${\cal L}_{1b}$ in a detection history is given by the
relation 
\[
{\cal L}_{1a}\widehat{U}_{0}\widehat{\rho }(R,\theta ,\phi )\widehat{U}%
_{0}^{\dagger }=\Gamma R^{2}g_{a}(\theta ,\phi )\widehat{U}_{0}\widehat{\rho 
}(R,\theta ,\phi )\widehat{U}_{0}^{\dagger }, 
\]
\begin{equation}
{\cal L}_{1b}\widehat{U}_{0}\widehat{\rho }(R,\theta ,\phi )\widehat{U}%
_{0}^{\dagger }=\Gamma R^{2}g_{b}(\theta ,\phi )\widehat{U}_{0}\widehat{\rho 
}(R,\theta ,\phi )\widehat{U}_{0}^{\dagger },  \label{jumpcoup}
\end{equation}
with 
\begin{equation}
g_{a}(\theta ,\phi )=\frac{1}{2}(1+\overrightarrow{u}\cdot \overrightarrow{u}%
_{a}),g_{b}(\theta ,\phi )=\frac{1}{2}(1+\overrightarrow{u}\cdot 
\overrightarrow{u}_{b}).  \label{gagb}
\end{equation}
Notice that these factors add up to $\Gamma R^{2}$. The contribution to the
density matrix arising from the history $\{t_{i},s_{i}\}$ is now easily
found in the form 
\[
\widehat{\rho }_{L}\left( \left\{ t_{i},s_{i}\right\} ,T\right) =\exp
[-R^{2}(1-e^{-\Gamma T})]\prod_{i=1}^{L}(\Gamma R^{2}e^{-\Gamma t_{i}}) 
\]
\begin{equation}
\times \int d\Omega f(\theta ,\phi )g_{a}^{n_{a}}(\theta ,\phi
)g_{b}^{n_{b}}(\theta ,\phi )\widehat{U}_{0}\widehat{\rho }(Re^{-\Gamma
T/2},\theta ,\phi )\widehat{U}_{0}^{\dagger },  \label{partialdens2}
\end{equation}
which looks quite similar as eq. (\ref{partialdens}). The probability
distribution for detection histories is given by the trace of (\ref
{partialdens2}), and the detection statistics can be obtained in the same
way as above. In analogy to eq. (\ref{pT}), the probability $%
p(n_{a},n_{b},T) $ that in the time interval $[0,T]$ there were $n_{a}$
detections in channel $a$, and $n_{b}$ in channel $b,$ irrespective of their
order, is now 
\[
p(n_{a},n_{b}T)=P_{L}(T)p_{L}(n_{a},n_{b}), 
\]
where, as before, $P_{L}(T)$ is the Poissonian distribution of the total
number $L=n_{a}+n_{b}$ of detections in the interval $[0,T]$. The factor $%
p_{L}(n_{a},n_{b})$, which represents the probability that the $L$
detections are partitioned over the two detectors as $(n_{a},n_{b})$, is 
\begin{equation}
p_{L}(n_{a},n_{b})=%
{L \choose n_{a}}%
F(n_{a},n_{b}),  \label{plab}
\end{equation}
with 
\begin{equation}
F(n_{a},n_{b})=\int d\Omega f(\theta ,\phi )g_{a}^{n_{a}}(\theta ,\phi
)g_{b}^{n_{b}}(\theta ,\phi ).  \label{Fab}
\end{equation}

As an example, we consider the case that before the coupling period the two
modes are fully decoupled, with equal population, so that the function $f$
is uniform over the equator of the sphere. The density matrix before
coupling has then the form (\ref{densfact}), with $\theta =\pi /2$. When
moreover the pulse duration is chosen such that $\delta \tau =\pi /2$, we
find $\overrightarrow{u}_{a}=-\widehat{y}$, $\overrightarrow{u}_{b}=\widehat{%
y}$, and the functions $g_{a}$ and $g_{b}$ at the equator are found as $%
g_{a}(\phi )=(1-\sin \phi )/2$, $g_{b}(\phi )=(1+\sin \phi )/2$. The
distribution $p_{L}(n_{a},n_{b})$ is now exactly the same as in the case of
an initally uniform phase distribution, with detectors are placed in the
output channel of a single $50\%-50\%$ beam splitter \cite{Nienhuis}, and we
recover the bunching distribution

\[
p_{L}(n_{a},n_{b})=\frac{1}{2^{2L}}%
{2n_{a} \choose n_{a}}%
{2n_{b} \choose n_{b}}%
. 
\]
The most probable history of $L$ detections is $(n_{a},n_{b})=(L,0)$ or $%
(0,L)$. This is understandable, since the relative geometry on the Bloch
sphere of the initial state and the detection operators is the same in both
cases. In the case of detections through the beam splitter, the distribution 
$f$ is initially uniform over the equator, and the detectors correspond in
the sense of eq. (\ref{trans}) to two opposite points on the equator. A
typical detection history then projects the phase distribution onto a narrow
peak located at either one of the detector unit vectors. In the case of
detectors attached directly to the two modes, the mode coupling prior to the
detections rotates the uniform distribution over the equator about the $x$%
-axis over an angle $\pi /2$, so that the initial distribution before the
detection series is uniform over the large circle in the $xz$-plane. The
detectors represented by the detection operators $\widehat{a}$ and $\widehat{%
b}$ correspond to the unit vectors $\pm \widehat{z}$, which again are two
point on opposite sides of the large circle representing the initial state.
However, the physical situation is quite different in the two cases. For the
initially uncoupled states and detection through the beam splitter, the
relative phase distribution starts out uniform, and it is converted into a
narrow distribution during a typical detection history in the output
channels of the beam splitter. For the initially coupled modes and the
detections without the beam splitter, the relative phase is initially rather
well-determined around $\phi =0$ and $\phi =\pi $. A typical detection
series now projects the state of the system onto the state with all
particles either in mode $A$ or in mode $B$, with an undetermined relative
phase. If at the end of the detection series a second pulsed coupling is
applied as described by the operator $\widehat{U}_{0}$, the final state
after this pulse has a well-determined relative phase. The net result of the
entire scheme of pulsed coupling, detection series and second pulse is the
same as the result of just a detection series through the beam splitter. In
this sense, the pulsed coupling can be viewed as a replacement of the beam
splitter.

\subsection{Continuous coupling between modes}

The situation is different when the coupling between the modes is present
continuously. Then in expression (\ref{L0}) for the coherent-evolution
operator, the Hamiltonian is given by eq. (\ref{Hcoup}). Since the
Hamiltonian commutes with the number operator $\widehat{N}$, the decay terms
are not affected the Hamiltonian evolution, and eq. (\ref{cohevol}) is
replaced by the modified form 
\begin{equation}
e^{{\cal L}_{0}T}\widehat{\rho }(R,\theta ,\phi )=\exp [-R^{2}(1-e^{-\Gamma
T})]\widehat{U}(T)\widehat{\rho }(Re^{-\Gamma T/2},\theta ,\phi )\widehat{U}%
^{\dagger }(T)  \label{UcohevolU}
\end{equation}
with $\widehat{U}(T)=\exp (-i\widehat{H}T/\hbar )=\exp (i\delta T\widehat{J}%
_{x}).$ The effect of the Hamiltonian on the density matrix for a detection
history $\left\{ t_{i},s_{i}\right\} $ can be expressed in the Heisenberg
picture, with the time-dependent detection operators 
\begin{equation}
\widehat{c}_{s}(t_{s})=\widehat{U}^{\dagger }(T)\widehat{c}_{s}\widehat{U}%
(T).  \label{UcU}
\end{equation}
Their action on the density matrix follows from eq. (\ref{jump}) when one
uses that $\widehat{c}_{a}(t)$ corresponds to the direction $\overrightarrow{%
u}_{a}(t)=-\widehat{y}\sin \delta t+\widehat{z}\cos \delta t$, and $\widehat{%
c}_{b}(t)$ to the opposite direction $\overrightarrow{u}_{b}(t)=\widehat{y}%
\sin \delta t-\widehat{z}\cos \delta t$. This gives 
\begin{equation}
\widehat{c}_{s}(t_{s})\widehat{\rho }(R,\theta ,\phi )\widehat{c}%
_{s}^{\dagger }(t_{s})=R^{2}g_{s}(\theta ,\phi ,t_{s})\widehat{\rho }%
(R,\theta ,\phi ),  \label{jump2}
\end{equation}
with $g_{s}(\theta ,\phi ,t)=(1+\overrightarrow{u}\cdot \overrightarrow{u}%
_{s}(t))/2$. The general expression (\ref{integmaster}) for the contribution
to the density matrix from a detection history $\left\{ t_{i},s_{i}\right\} $
with the initial state (\ref{inital}), is found as 
\[
\widehat{\rho }_{L}\left( \left\{ t_{i},s_{i}\right\} ,T\right) =\exp
[-R^{2}(1-e^{-\Gamma T})]\prod_{i=1}^{L}(\Gamma R^{2}e^{-\Gamma t_{i}}) 
\]
\begin{equation}
\times \int d\Omega f(\theta ,\phi )\prod_{i=1}^{L}[g_{s_{i}}(\theta ,\phi
,t_{i})]\widehat{U}(T)\widehat{\rho }(Re^{-\Gamma T/2},\theta ,\phi )%
\widehat{U}^{\dagger }(T).  \label{rhocoup}
\end{equation}
Each detection $s$ leads to a multiplication of the distribution function
over the Bloch sphere by a factor $g_{s}(\theta ,\phi ,t)$ that now depends
on the detection time. This time dependence corresponds to a rotation of the
direction $\overrightarrow{u}_{s}$ in the $yz$-plane.

For the initial state of two decoupled modes, with a uniform distribution of
the phase, the function $f$ is uniform over the equator of the Bloch sphere.
A detection at time $t$ of a particle emitted by mode $A$ or $B$ then
multiplies the distribution over the relative phase $\phi $ by the factor $%
g_{a}(\phi )=(1-\sin \delta t\sin \phi )/2$ , or $g_{b}(\phi )=(1+\sin
\delta t\sin \phi )/2$. These functions have their maximum value for $\phi
=3\pi /2$ or $\phi =\pi /2$. Strictly speaking, this distribution describes
the state of the system in the Heisenberg picture, where it is not affected
by continuous evolution, but only by the quantum jumps that describe the
effect of detections. The evolution of the phase distribution during a
typical detection history is conceptually simple. The total decay rate,
summed over both detectors, is autonomous, and has the time dependent rate $%
\Gamma R^{2}\exp (-\Gamma t)$. The branching over the two detectors $a$ and $%
b$ is determined by the expectation value of $g_{a}(\phi )$ and $g_{b}(\phi
) $, which has a contrast that oscillates in time at the coupling frequency $%
\delta $, as a result of the mode coupling. The effect of a detection on the
phase distribution is a multiplication with the same factor $(1\mp \sin
\delta t\sin \phi )/2$, for detector $a$ and $b$. This will eventually lead
to convergence to the phase distribution to a single peak at a value where
either one of the factors $g_{s}$ is maximal, hence $\phi =\pi /2$ or $\phi
=3\pi /2$. The convergence to these peaked distributions is slower than in
the case of a detections through a single beam splitter, as a result of the
oscillations of the contrast in the functions $g_{s}(t)$. In Fig. 4 we plot
a set of typical phase distributions after $L=10$ detections. The instants
of detection are randomly selected, and the most probable dteection channel
at that instant is chosen. The different curves correspond to a different
selection of the instants of detection. As seen in Fig. 4, after each such
history, the distribution over $\phi $ is a peak centered either at $\pi /2$
or at $3\pi /2$.

\subsection{Coupling and energy shift}

An energy difference $\hbar \varepsilon $ between the two modes in addition
to the effect of tunneling is described by the Hamiltonian 
\begin{equation}
\widehat{H}=-\hbar \delta \widehat{J}_{x}+\hbar \varepsilon \widehat{J}_{z},
\label{Hshift}
\end{equation}
which replaces (\ref{Hcoup}). The angular-momentum operators are defined in
eq. (\ref{Schwinger}). We consider the same detection scheme used in the
preceding subsection. The energy difference modifies the detection
statistics and the phase distribution following a representative detection
history. On the Bloch sphere, the modified evolution operator $\widehat{U}%
(t) $ is represented by a rotation in the positive direction around the axis 
$\varepsilon \widehat{z}-\delta \widehat{x},$ over an angle $\Omega t$, with 
$\Omega =\sqrt{\varepsilon ^{2}+\delta ^{2}}$. Equations (\ref{UcohevolU})
for the density matrix after a detection history and (\ref{UcU}) for the
detection operators in the Heisenberg representation $\widehat{c}_{s}(t)$
remain valid. The detection operators are represented by points $%
\overrightarrow{u}_{s}$ on the sphere that are reached from the poles when
the opposite rotation is applied. Since the rotation axis does not lie in
the equator plane, the azimuthal angle varies continuously with time, and
the relative phase is no longer projected preferentially onto the same
value. These unit vectors are found in the form 
\[
\overrightarrow{u}_{a}(t)=-\overrightarrow{u}_{b}(t)=\frac{\varepsilon
\delta }{\Omega ^{2}}(\cos \Omega t-1)\widehat{x}-\frac{\delta }{\Omega }%
\sin \Omega t\widehat{y}+\left( \frac{\delta ^{2}}{\Omega ^{2}}\cos \Omega t+%
\frac{\varepsilon ^{2}}{\Omega ^{2}}\right) \widehat{z}. 
\]
They determine the factors $g_{s}(\theta ,\phi ,t)=(1+\overrightarrow{u}%
\cdot \overrightarrow{u}_{s}(t))/2$ that multiply the distribution over the
sphere when a particle emitted by mode $A$ or $B$ is detected.

As above, we consider the case of an initially factorized state, which is
represented by a uniform distribution over the equator of the Bloch sphere.
When a particle from mode $A$ or $B$ is detected, the distribution over $%
\phi $ is multiplied by 
\[
g_{a}(\phi )=\frac{1}{2}\left( 1+\frac{\varepsilon \delta }{\Omega ^{2}}\cos
\phi (\cos \Omega t-1)-\frac{\delta }{\Omega }\sin \phi \sin \Omega t\right)
, 
\]
\[
g_{b}(\phi )=\frac{1}{2}\left( 1-\frac{\varepsilon \delta }{\Omega ^{2}}\cos
\phi (\cos \Omega t-1)+\frac{\delta }{\Omega }\sin \phi \sin \Omega t\right)
. 
\]
The maximum of these functions no longer coincide with the maximum of $\pm
\sin \phi $, as is the case when $\varepsilon =0$.

In Fig. 5 the resulting phase distributions are shown after a number of
typical detection histories, each consisting of $10$ detections, for $%
\varepsilon /\delta =1/4$. The prescription of the calculation is the same
as used in Fig. 4. Now not only the width of the peak, but also their
position varies for different selections of the detection times. This can be
explained from the variation in the position where the maximum of $%
g_{s}(\phi ,t)$ occurs.

\section{Linear and circular chains of modes}

The dynamics of a coupled chain of condensates in an optical lattice has
been explored, with emphasis on the difference between a linear and a
circular chain \cite{Tsukada}. The coupling was due to tunneling between
neighboring modes. One expect anaologous differences in the situation
considered in this paper, where the phase relation between neighboring modes
arises by spontaneous symmetry breaking from the observation of emitted
bosons interfering through a beam splitters. This raises the question of the
transitivity of the relative phase. When the relative phase between two
modes $A$ and $B$ is well-determined, and the same holds for the relative
phase between two modes $B$ and $C$, then one expects the phase between $C$
and $A$ should also be fixed. On the other hand, when this latter phase is
also selected by direct interaction, one may expect different dynamics
depending on whether the two paths of phase determination converge to the
same result or not. In the present section we compare the phase dynamics on
a linear and a circular chain of modes.

\subsection{Linear chain of modes}

We consider a linear chain of modes, as sketched in Fig. 6. As initial state
we take the uncorrelated state given by the factorized density matrix 
\begin{equation}
\widehat{\rho }(0)=\prod_{s}\widehat{\rho }_{s}=\ldots \widehat{\rho }%
_{s-1}\otimes \widehat{\rho }_{s}\otimes \widehat{\rho }_{s+1}\ldots ,
\label{initline}
\end{equation}
where the density matrix $\widehat{\rho }_{s}$ of each mode $s$ has the form
(\ref{poissonian}) with a uniform phase $\phi _{s}$. Beam splitters are
mixing the bosons emitted from neighboring modes $s$ and $s+1$, with
orthogonal detection operators in the output channels 
\begin{equation}
\widehat{d}_{s\pm }=\frac{1}{\sqrt{2}}(\widehat{a}_{s}\pm e^{-i\xi _{s}}%
\widehat{a}_{s+1}).  \label{operline}
\end{equation}
with $\widehat{a}_{i}$ the annihilation operator of mode $i$. The evolution
is described by the master equation (\ref{master}), with 
\begin{equation}
{\cal L}_{0}\widehat{\rho }=-\sum_{s}\frac{\Gamma }{2}(\widehat{a}%
_{s}^{\dagger }\widehat{a}_{s}\widehat{\rho }+\widehat{\rho }\widehat{a}%
_{s}^{\dagger }\widehat{a}_{s}),{\cal L}_{1}=\sum_{s}({\cal L}_{1s+}+{\cal L}%
_{1s-}),  \label{Lline}
\end{equation}
where the contribution to ${\cal L}_{1}$ corresponding to the detection
channels $s_{\pm }$ are specified by 
\begin{equation}
{\cal L}_{1s\pm }=\frac{\Gamma }{2}\widehat{d}_{s\pm }\widehat{\rho }%
\widehat{d}_{s\pm }^{\dagger }.  \label{L1line}
\end{equation}
Physically it is obvious that the detection statistics over the output
channels of each beam splitter is identical to the statistics for each of
the two beam splitters in Sec. III, since each mode emits into two input
channels with equal rate. The density matrix corresponding to a given
detection history with $n_{s}\det $ections in channel $s_{+}$, and $m_{s}$
detections in the channel $s_{-}$ is easily written down by using that a
detection in channel $s_{+}$ gives a factor $\cos ^{2}((\Phi _{s}-\xi
_{s})/2)$, and a detection in channel $s_{-}$ a factor $\sin ^{2}((\Phi
_{s}-\xi _{s})/2)$. After each detection history, the distribution over the
phases $\phi _{s}$ of all modes factorizes into a product of distributions
for each relative phase $\Phi _{s}\equiv \phi _{s}-\phi _{s+1}$ between
neighbors. After $n_{s}\det $ections in channel $s_{+}$, and $m_{s}$
detections in the channel $s_{-}$, the distribution over the relative phase $%
\phi _{s}-\phi _{s+1}$ is proportional to $\cos ^{2n_{s}}((\Phi _{s}-\xi
_{s})/2)\sin ^{2m_{s}}((\Phi _{s}-\xi _{s})/2)$, and the distribution over
the phases is proportional to the product 
\begin{equation}
\prod_{s}[\cos ^{2n_{s}}(\frac{\Phi _{s}-\xi _{s}}{2})\sin ^{2m_{s}}(\frac{%
\Phi _{s}-\xi _{s}}{2})].  \label{phasedist}
\end{equation}
Because of this factorization, the detection statistics for the pair of
output channels of each beam splitter is uncorrelated to the other
detections. The total number $M_{s}$ of detections in the time interval $%
[0,T]$ on the two output channels of a single beam splitter is Poissonian
with average value $r^{2}[1-\exp (-\Gamma T)]$, and the probability
distribution of the $M_{s}$ detections over the two detectors is identical
to the distribution (\ref{uniform}) \cite{Nienhuis}. Therefore, the most
probable histories with $M_{s}$ detections on this $s$th beam splitter are
given as $(n_{s},m_{s})=(M_{s},0)$ and $(0,M_{s})$. The relative phase $\Phi
_{s}$ between modes $s$ and $s+1$ converges to a single peak located at $\xi
_{s}$ or $\xi _{s}+\pi $, for each value of $s$. This also determines in a
unique and unambiguous way the relative phase between any pair of modes.
Hence for a linear chain of modes, the relative phase between two neighbors
converges to one out of two possible values, in precisely the same way as it
occurs for two modes and a single beam splitter. Spontaneous symmetry
breaking occurs independently for each neighboring pair.

\subsection{Circular chain of modes}

Now we consider a series of $K$ modes, coupled by beam splitters, and
arranged into a circular chain. For $K=3$, the scheme is presented in Fig.
7. Equations (\ref{initline})-(\ref{Lline}) still hold, with the index $s$
running from $1$ to $K$. The relative phases $\Phi _{s}$ and the detection
operators $\widehat{d}_{s\pm }$ are defined as above for $s=1$, $2$,$\ldots $%
, $K-1$, while we denote $\Phi _{K}=\phi _{K}-\phi _{1}$, $\widehat{d}_{K\pm
}=(\widehat{a}_{K}\pm e^{-i\xi _{K}}\widehat{a}_{1})/\sqrt{2}$. The number
of beam splitters is now equal to the number of modes. On the other hand,
since 
\begin{equation}
\sum_{s=1}^{K}\Phi _{s}=0,  \label{PhaseRelation}
\end{equation}
the $K$ modes have only $K-1$ independent relative phases $\Phi _{s}$, which
makes the detection system overdetermined. This is the main difference with
the case of the linear chain. Detections on the $s$th beam splitter tend to
drive the relative phase $\Phi _{s}$ to the value $\xi _{s}$ or $\xi
_{s}+\pi $. However, these values are consistent only when the values of all 
$\xi _{s}$ add up to a multiple of $\pi $. The probability $%
p(\{n_{s},m_{s}\},T)$ of a specified number of detections by each detector
in the time interval $[0,T]$ factorizes as in eq. (\ref{pT}) in a Poisson
distribution for the total number $L$ of detections, with the mean value $%
Kr^{2}(1-e^{-\Gamma T})$, and the probability $p_{L}(\{n_{s},m_{s}\})$ that
the $L$ detections are distributed over the detectors according to the
indicated partition. This latter distribution can be specified in analogy to
(\ref{pL}) by 
\begin{equation}
p_{L}(\{n_{s},m_{s}\})=\frac{L!}{\prod_{s}(n_{s}!m_{s}!)}F(\{n_{s},m_{s}\})
\label{pLcirc}
\end{equation}
with 
\begin{equation}
F(\{n_{s},m_{s}\})=\left( \frac{1}{2\pi }\right) ^{K}\int d\phi _{1}d\phi
_{2}\ldots d\phi _{K}\prod_{s=1}^{K}\left[ \cos ^{2n_{s}}(\frac{\Phi
_{s}-\xi _{s}}{2})\sin ^{2m_{s}}(\frac{\Phi _{s}-\xi _{s}}{2})\right] .
\label{Fcoup}
\end{equation}
After a detection history with $n_{s}\det $ections in channel $s_{+}$, and $%
m_{s}$ detections in the channel $s_{-}$, the distribution over the relative
phase is still proportional to (\ref{phasedist}). However, because of the
relation (\ref{PhaseRelation}), the relative phases are no longer
independent, and the detection statistics of the output channels of the
different beam splitters become correlated.

The most probable histories can now be found by similar considerations as we
used above in Sec. IIID. For a total number of $L=K\times M$ detections, one
might expect that the same number ($M$) of particles reaches each beam
splitter, with the partition $(n_{s},m_{s})=(M,0)$ or $(0,M)$ for all of the 
$K$ beam splitters. This would indicate that the corresponding relative
phases probed by these beam splitters will have converged to the value $\xi
_{s}$ or $\xi _{s}=\pi $. However, in general this can only be true for all
relative phases except one, because of the phase relation (\ref
{PhaseRelation}). Assume that this excepted relative phase has the index $%
s_{0}$. As a result of this relation, the value of the last relative phase $%
\Phi _{s_{0}}$ is thereby also fixed. The distribution over the two output
channels $s_{0+}$ and $s_{0-}$ will then be binomial, and the most probable
partition is given by $(n_{s_{0}},m_{s_{0}})$=$(M\cos ^{2}((\Phi
_{s_{0}}-\xi _{s_{0}})/2),M\sin ^{2}((\Phi _{s_{0}}-\xi _{s_{0}})/2)$. For
symmetry reasons, each beam splitter has the same probability to end up in
such a binomial distribution rather than a bunching one. The situation can
be summarized by stating that in addition to the local spontaneous symmetry
breaking for each beam splitter, also a global symmetry breaking occurs, by
which the relative phase between two neighbors is not determined by the
setting of their own shared beam splitter, but by the settings of all the
other ones.

We backed up this conclusion by a numerical calculation in the case of a
three-mode ring, as sketched in Fig. 7. The settings of the beam splitters
are given by $\xi _{1}=\xi _{2}=0$, $\xi _{3}=\pi /2$. After $30$
detections, one of the partitions with the highest probability was found to
be $(n_{1},m_{1})=(5,5)$, $(n_{2},m_{2})=(10,0)$, $(n_{3},m_{3})=(10,0)$. As
one would expect from symmetry considerations, other partitions with the
same maximal probability are found by swapping $n_{s}$ and $m_{s}$ for each
beam splitter $s$, and also by a permutation of the three indices $1$, $2$
and $3$.

\section{Discussion and conclusions}

The absolute phase of a single-mode or multimode bosonic system is fully
undetermined when the state of the system is diagonal in the total particle
number. For bosonic atoms, this must be the case, since states with
different particle numbers do not superpose. For a two-mode system we use
the Schwinger representation with fictitious angular momentum operators to
take advantage of the underlying $SU(2)$ symmetry of the state space. This
allows us to represent the density matrix of the two-mode system with an
undetermined absolute phase and a Poissonian distribution of the total
number of particles as an integral over the Bloch sphere of the fictitious
angular momentum. The representation is given in eq. (\ref{dens}), where $%
f\left( \theta ,\phi \right) $ is the distribution function over the sphere.
It may be viewed as the Glauber-Sudarshan $P$ function restricted to the
sphere. The azimuthal angle $\phi $ is the relative phase, whereas the polar
angle $\theta $ measures the ratio of the average number of particles in $A$
and $B$, with equal populations represented by points on the equator, and
the poles representing states with all particles in one mode. The merit of
these states with Poissonian distribution of the total particle number is
that the overall decay of the modes factors out, and the detection
statistics is the product of time-dependent probabilities for the total
number of detections, and time-independent distributions for the partitions
over the various detection channels. The effect of a detection is described
by the action of an annihilation operator, which also corresponds to a point
on the sphere. This is equivalent to the multiplication of the distribution
function $f\left( \theta ,\phi \right) $ by a factor that depends only on
the distance over the sphere between the points $\left( \theta ,\phi \right) 
$ and the detection point. This allows exact expressions, both for the
detection statistics, and for the conditional density matrix of the system
for a given detection history. It also implies that identical detection
statistics arises for different choices of the distribution $f$ and the
detection points on the Bloch sphere, provided that the setup has the same
relative geometry on the sphere. This can correspond to quite different
experimental setups, since the effect of detection through a beam splitter
can be produced by a pulsed tunneling coupling between the modes.

In the case that the modes are constantly coupled by tunneling, and in the
presence of an energy difference between the modes, the phase distribution
still becomes non-uniform by the detecting particles emitted by the two
modes. However, since the preferred phase imposed by the detections is not
the same for all detections in this case, the maximum in the phase
distribution will continue to vary in position even after many detections.
The convergence of the phase will be perturbed more strongly when
interparticle interactions are important during a detection history.

We treat explicitly the case of two modes which both emit particles in an
input channel of two different beam splitters. When the settings of the beam
splitters are different, they can drive the relative phase of the modes to
values which are conflicting. Such a situation of conflicting phase values
occurs for any number of modes which are coupled by beam splitters, and
arranged in a circular chain. Our model shows that in these cases the most
probable detection histories lead for each pair of neighboring modes to a
relative phase converging with equal probability to one of the conflicting
values. The partition of the detection over the channels is a signature of
the location of the peak in the phase distribution. Such a conflict does not
arise for a linear chain of modes coupled by a beam splitter. A common
feature of these various cases is that an initially factorized state of
several modes builds up a specific value of all relative phases by only
detecting their decay products in interference. In principle, this means
that the modes become entangled, even though they have never been in direct
contact.

\acknowledgments
This work is part of the research program of the ``Stichting voor
Fundamenteel Onderzoek der Materie'' (FOM).

\begin{figure}[tbp]
\centerline{\psfig{figure=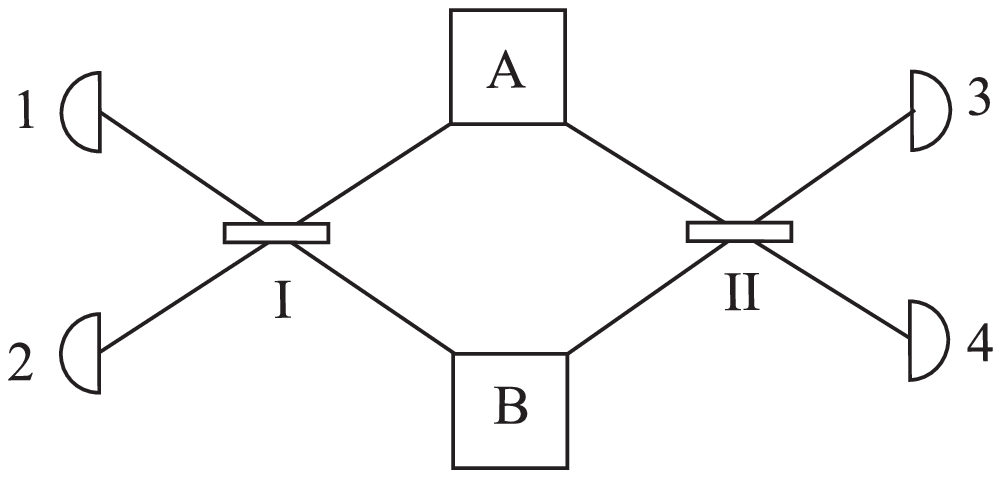}}
\caption{ Sketch of setup with two decaying modes, where each mode emits
particles into the input port of two different beam splitters.}
\end{figure}
\newpage

\begin{figure}[tbp]
\centerline{\psfig{figure=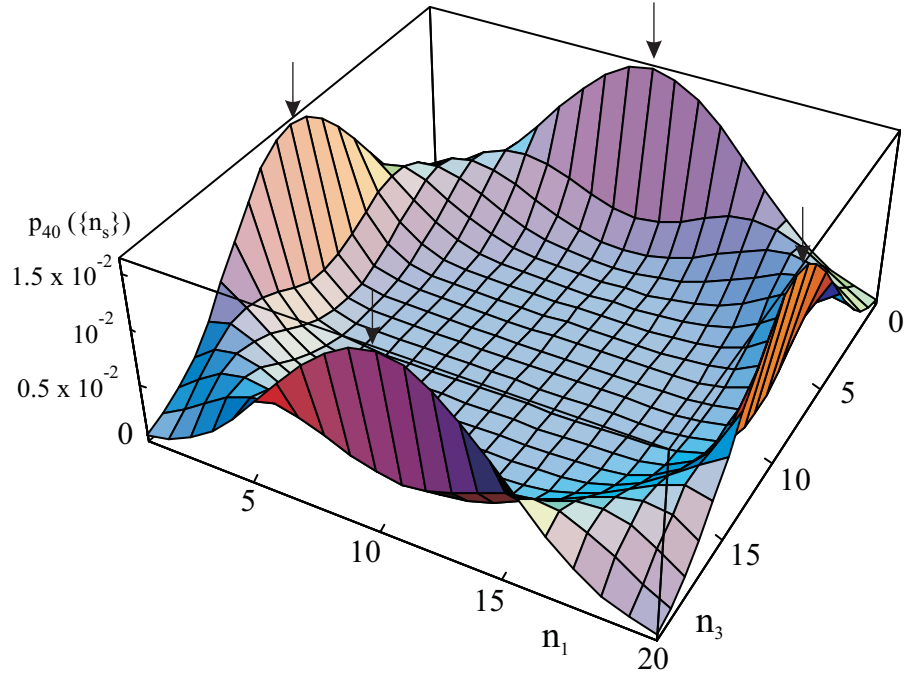}}
\caption{ The probability distribution $p_L(\{n_s\})$ versus $n_1$ and $n_3$
for equal number of detections through both beam splitters. Here $L=40$ and $%
\xi =\pi /2$. The most probable histories are marked.}
\end{figure}
\newpage

\begin{figure}[tbp]
\centerline{\psfig{figure=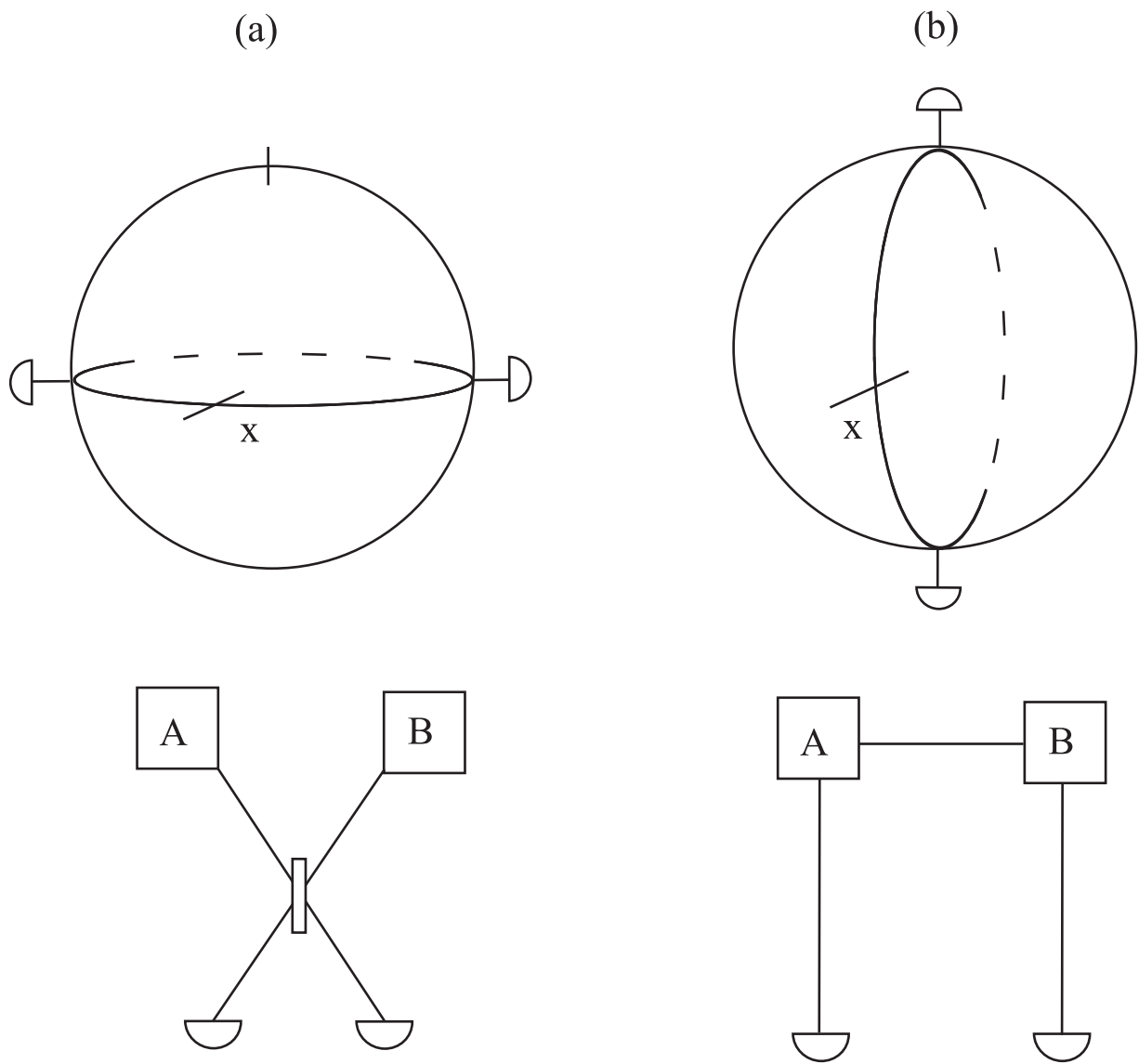}}
\caption{(a) emitted particles are detected directly, without the use of
beam splitters; (b) emitted particles re detected through a beam splitter.
For each case, the position of the detectors on the Bloch sphere, and the
distribution of the state before detection are also shown.}
\end{figure}
\newpage

\begin{figure}[tbp]
\centerline{\psfig{figure=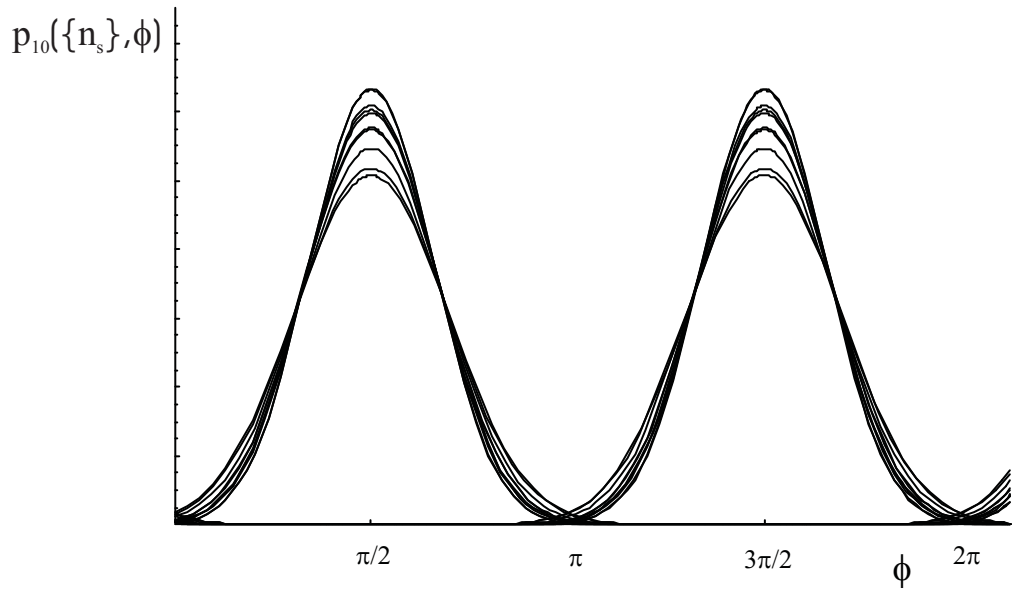}}
\caption{ Relative phase distributions for two coupled modes after $L=10$
detections. Each curve of the ten curves corresponds to a different
realization of the randomly selected detection times.}
\end{figure}
\newpage

\begin{figure}[tbp]
\centerline{\psfig{figure=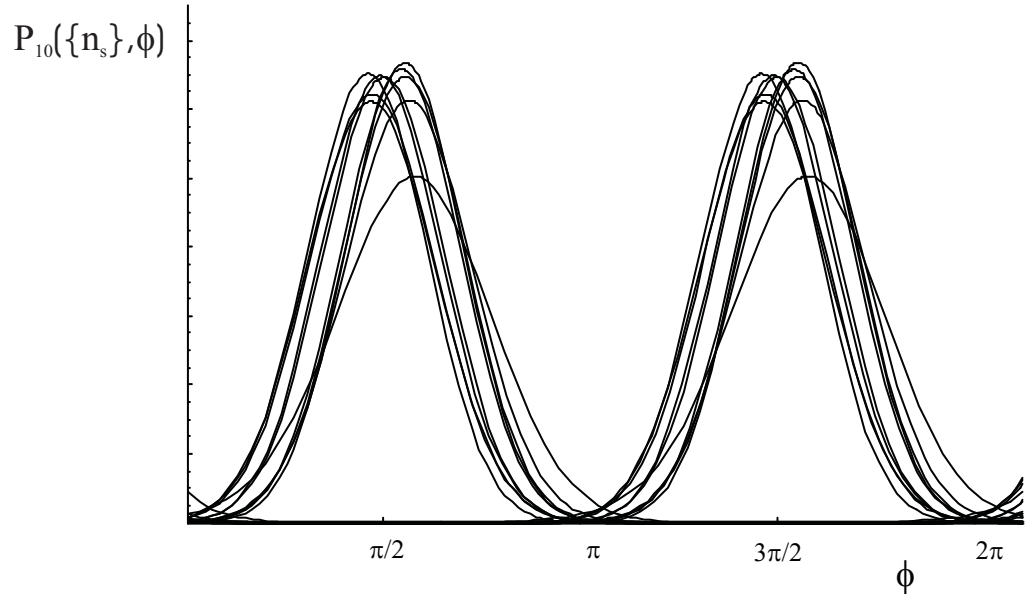}}
\caption{ Same as fig. 4, now for two coupled modes at different energy.
Coupling strength and energy splitting specified by $\varepsilon /\delta =1/4
$.}
\end{figure}
\newpage

\begin{figure}[tbp]
\centerline{\psfig{figure=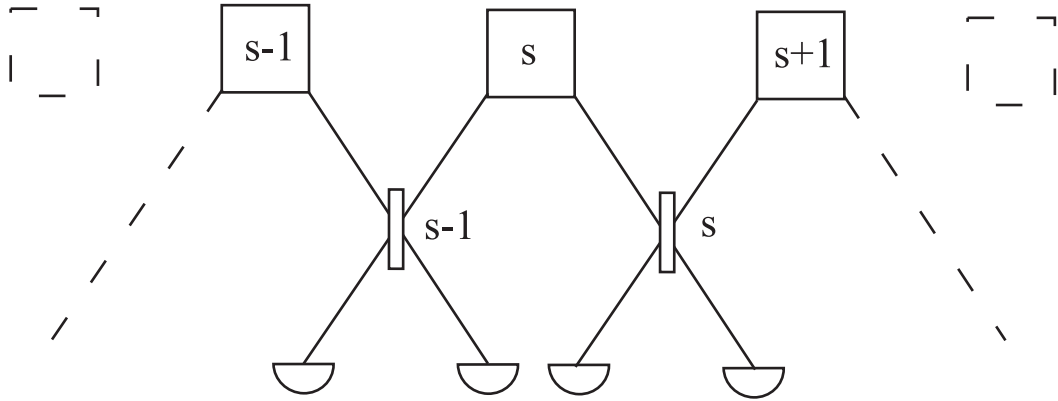}}
\caption{Sketch of setup with a linear chain of modes}
\end{figure}
\newpage

\begin{figure}[tbp]
\centerline{\psfig{figure=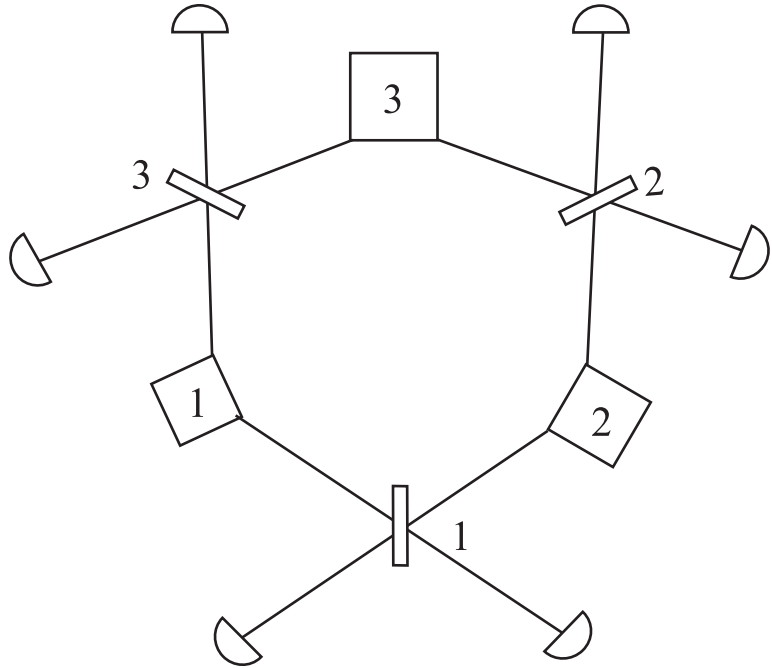}}
\caption{Sketch of setup with 3 modes arranged on circular chain.}
\end{figure}
\newpage

\end{document}